\documentclass{ws-procs961x669}            

\usepackage{hyperref}
\usepackage[table]{xcolor}

\begin{document}
\title{Joint Analysis Method on Gravitational Waves and Low-Energy Neutrinos to Detect Core-Collapse Supernovae
}

\author{O. Halim$^*$}
\address{Istituto Nazionale di Fisica Nucleare (INFN) sez. di Trieste, Italy\\
$^*$E-mail: odysse.halim@ts.infn.it\\
}

\author{C. Casentini}
\address{Istituto Nazionale di Astrofisica - Istituto di Astrofisica e Planetologia Spaziali (INAF -IAPS), Rome, Italy,\\
}

\author{M. Drago}
\address{Universit\`a di Roma  La Sapienza, I-00185 Roma, Italy,\\
INFN, Sezione di Roma, I-00185 Roma, Italy}

\author{V. Fafone}
\address{University of Rome Tor Vergata, Rome, Italy,\\
INFN sez. di Roma Tor Vergata, Rome, Italy,}

\author{K. Scholberg}
\address{Department of Physics, Duke University, Durham, NC, USA}

\author{C. F. Vigorito}
\address{University of Turin, Italy,\\
INFN sez. di Torino, Italy,}

\author{G. Pagliaroli}
\address{Gran Sasso Science Institute (GSSI), L'Aquila, Italy,\\
INFN sez. di LNGS, Assergi, Italy}

\begin{abstract}
Core-collapse supernovae produce copious low-energy neutrinos and are also predicted to radiate gravitational waves. These two messengers can give us information regarding the explosion mechanism. The gravitational wave detection from these events are still elusive even with the already advanced detectors. Here we give a concise and timely introduction to a new method that combines triggers from GW and neutrino observatories; more details shall be given in a forthcoming paper \cite{Halim2021}.


\end{abstract}

\keywords{multimessenger, supernova, core-collapse, low-energy neutrino, gravitational wave.}

\bodymatter

\section{Introduction}\label{sec:intro}
Core-collapse supernovae (CCSNe) are expected to produce multimessenger signals such as neutrinos, gravitational waves (GWs), as well as multi-wavelength electromagnetic waves \cite{pagliaroli_PRL,leonor}. Low-energy neutrinos (LENs) are expected to be produced by these events, with the average energy $\sim 10$ MeV and these LENs are from the majority of the total energy budget ($\sim 10^{53}$ erg) of the CCSNe.



LENs from a CCSN have been successfully detected with the observation of SN1987A in Large Magellanic Cloud by Kamiokande-II \cite{hirata1987}, IMB \cite{Bionta1987}, and Baksan \cite{alexeyev}. Today, there are several LEN detectors waiting for these astrophysical events such as Super-Kamiokande \cite{superK} (Super-K), LVD \cite{lvd_det}, KamLAND \cite{kamland}, and IceCube \cite{icecube} with the horizon up to the edge of our galaxy and beyond. These detectors are in fact in a multi-detector collaboration for this effort to produce low-latency alerts under SuperNova Early Warning System (SNEWS) \cite{Antonioli2004,Al_Kharusi_2021}.

Moreover, the era of multimessenger astronomy involving GWs has just begun with the detection of the binary neutron star merger \cite{Abbott2017}. Recently, GW search has been in the phase of O3 data taking, which is done by LIGO \cite{ligo} detectors (4-km arm in Hanford and Livingston USA), Virgo \cite{virgo} detector (3-km arm in Cascina Italy), and KAGRA \cite{KAGRA} (3-km arm in Gifu Prefecture, Japan in the latest period of O3). In total, the GW science data taking has been done for 3 observing runs (O1, O2, and O3) from 2015 to 2020, and currently the detectors undergo improvements in order to be more sensitive for O4 (expected in August 2022). CCSNe are expected to produce GW signals with broad physical processes \cite{Ott_2009,Abdikamalov:2020jzn,powell2020,Szczepanczyk}. Detecting GWs from these sources will enable us to study the physical processes. As we will see later on, the detection capability of GWs could be improved by multimessenger search involving LENs.

\begin{figure}[!ht]
\centering
\includegraphics[width=.7\linewidth]{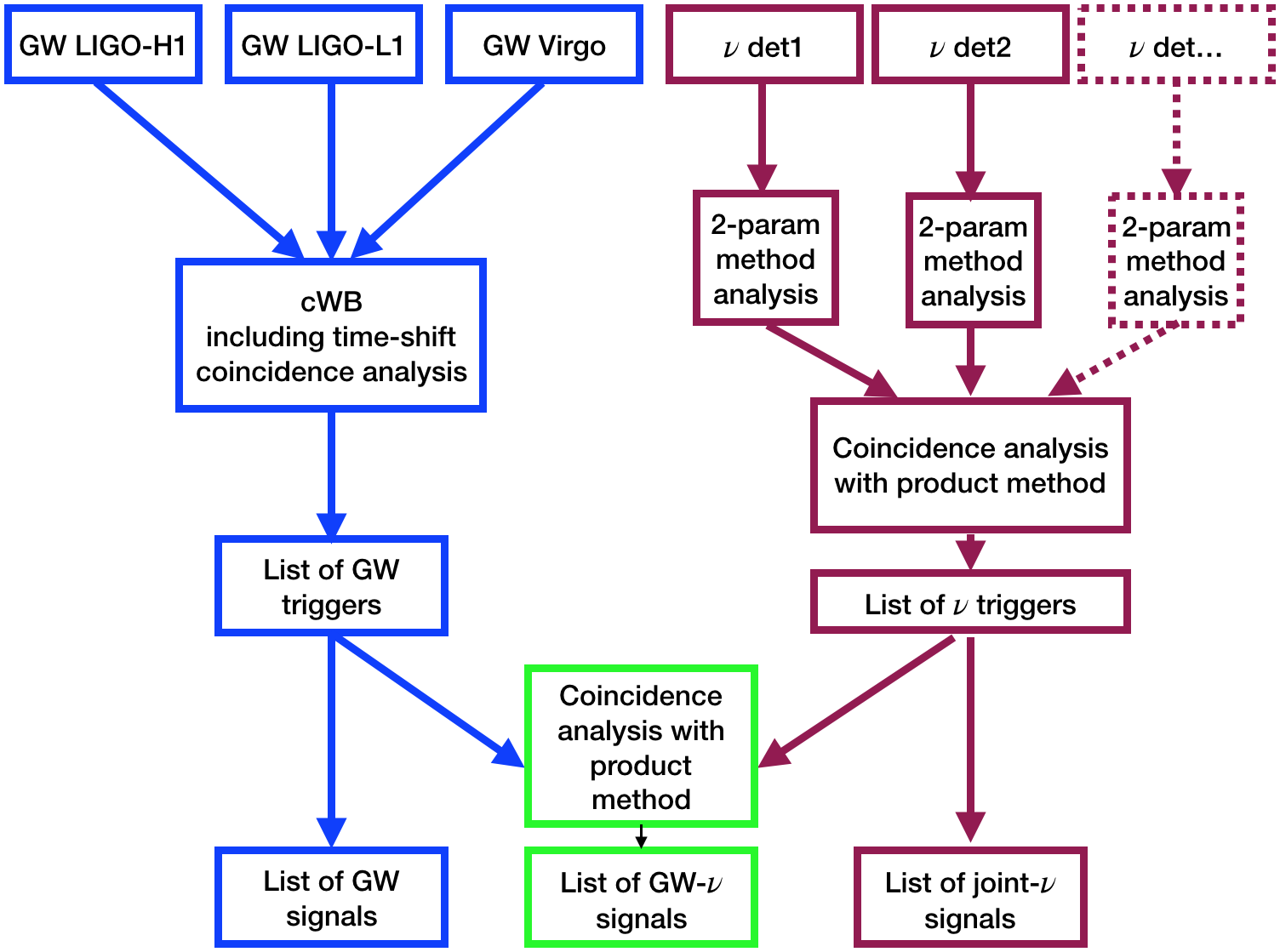}
\caption{The schematic view of the multimessenger GW-LEN strategy.}
\label{fig:GWnu_scheme}
\end{figure}

Here, we provide a timely description of a strategy to combine GWs and LENs for a multimessenger search following the chart in Fig.~\ref{fig:GWnu_scheme}. More details discussion will be given in a forthcoming paper \cite{Halim2021}, which has just been accepted by JCAP. This method is based on our previous works \cite{halimtesi,halim2019}. We construct a time-coincident strategy and test it with the simulated signals for both GW and LEN data. 



In our strategy, we use the \texttt{coherent WaveBurst} (cWB) pipeline \cite{Klimenko_2004,Klimenko2008,dragotesi,Necula_2012} for GW data analysis from simulations. This is a model-agnostic algorithm pipeline that is used to search GWs from CCSNe \cite{em_gw_ccsne}. Moreover, we also simulate LEN signals arriving in several neutrino detectors and analyse them\footnote{Note that we employ no detailed neutrino-detector simulation.}. At this stage, we use a new approach (introduced in \cite{halim2019}) for neutrino analysis in order to exploit the temporal behaviour of CCSNe. In the next step, we perform a temporal-coincidence analysis between the two messengers. Different messenger data are analysed separately and then combined together for possible GW-LEN signals. This could be interesting for online networks such as SNEWS or offline search for sub-threshold signals.


This paper is organised as follows. In Sec.~\ref{sec:sources}, we discuss the emission models for each messenger. Then, in Sec.~\ref{sec:science}, the data and analysis by our strategy will be presented. In the end, we show the implementation of our strategy in Sec.~\ref{sec:results}.



\section{Messengers from Core-collapse Supernovae}\label{sec:sources}

A CCSN is expected to produce a $O(10\,\mathrm{ms}-1\,\mathrm{s})$ GW burst and an impulsive $O(10-30\,\mathrm{s})$ emission of  $O(10\mathrm{\,MeV})$ LENs \cite{jankarev}\footnote{These signals could also be emitted by ``failed'' SNe \cite{OConnor:2010moj}.}, which highly depend on the CCSN initial conditions: progenitor mass, rotation, etc. Thus, it would be ideal to perform a simultaneous GW-LEN analysis from \textit{the same numerical simulation}. Unfortunately, no simultaneous GW-LEN numerical simulations have provided successful CCSN explosion, or if there are, the simulations can only provide both signals for the first half of a second, until the explosion. Meanwhile, the LEN emission duration is expected to last about 10 seconds. Therefore, we combine the currently available GW signals and LEN signals from different simulations with similar progenitor masses, and handle them as if they are from the simultaneous simulation.


\subsection{Gravitational Wave Emission}\label{sec:GWemission}

We provide the detail of our GW waveforms in Tab.~\ref{tab:gw_models}. We use the GW signals from the 3D neutrino-radiation hydrodynamics simulations of Radice {\it et al.} \cite{radice} (named as ``Rad'') with the various zero age main sequence masses in order to have the successful explosions from the low-mass and the failed explosions from the high-mass. Besides, we also take into account models with rapid rotation and high magnetic field which produce much stronger GWs. In this case, we take GW waveforms from two simulations: the Dimmelmeier \cite{dimmelmeier2008} (``Dim'') and the Scheidegger \cite{Scheidegger:2010en} (``Sch''). The stellar progenitors of Dim and Sch have strong rotation and magnetic field. These constraints make the models less favourable to happen than the neutrino-radiation mechanism \cite{em_gw_ccsne,jankarev,woosley}. Nevertheless, we would not rule out any models since we have not yet detected any CCSN GWs. Thus, we try to cover as broad as possible the uncertainty band from theoretical predictions: the lower limit is from the Rad model, and the upper limit is from the Dim and the Sch models.





\begin{table}
\tbl{Waveforms from the considered CCSN simulations. In the columns: emission type and reference, waveform identifier, waveform abbreviation for this manuscript, progenitor mass, angle-averaged root-sum-squared strain $h_\mathrm{rss}$, frequency when the GW energy spectrum peaks, and emitted GW energy.}
{\begin{tabular}{@{}ccccccc@{}}
\toprule
Waveform & Waveform & Abbr. & Mass  & $h_\mathrm{rss}\,@10\, \mathrm{kpc}$ & $f_\mathrm{peak}$  & $E_\mathrm{GW}$\\
    Family & Identifier & & $M_\odot$ &  $\mathrm{\left[10^{-22}\,\frac{1}{\sqrt{Hz}}\right]}$ & $\mathrm{[Hz]}$ & $[10^{-9}\,M_\odot c^2]$  \\\colrule
    Radice \protect\cite{radice}                  & s25 & Rad25 & {25} & 0.141 & 1132 & 28 \\
      3D simulation; & s13 & Rad13 & {13}   &  0.061 & 1364  & 5.9  \\ 
        $h_+$ \& $h_\times$; (Rad)            & s9 & Rad9 & 9  & 0.031 & 460 & 0.16  \\
    \hline
    Dimmelmeier \protect\cite{dimmelmeier2008}      & dim1-s15A2O05ls & Dim1 & 15 & 1.052 & 770 & 7.685 \\
      2D simulation; & dim2-s15A2O09ls & Dim2 & 15 & 1.803 & 754 & 27.880\\
       $h_+$ only; (Dim)          & dim3-s15A3O15ls & Dim3 & 15 & 2.690  &  237 &  1.380 \\
     \hline
     
     Scheidegger \protect\cite{Scheidegger:2010en} & sch1-R1E1CA$_L$ & Sch1 & 15  & 0.129  & 1155  & 0.104  \\
    3D simulation;   & sch2-R3E1AC$_L$ & Sch2 & 15  &  5.144 & 466  & 214  \\
      $h_+$ \& $h_\times$; (Sch)         & sch3-R4E1FC$_L$ & Sch3 & 15 &   5.796  & 698  &  342 \\\botrule
\end{tabular}}
\label{tab:gw_models}
\end{table}


\subsection{Low-energy Neutrino Emission}\label{sec:NUemission}


Similar to the previous emission, we consider two models for the LEN emissions. First, we take the signals from the numerical simulations of H{\"u}depohl obtained for a progenitor of $11.2 M_\odot$; without the collective oscillations \cite{hudepohl}, with the time-dependent neutrino luminosities, and average energies. The simulation provides the first 7.5 seconds of the neutrino emission with the analytical extension in order to reach 10 seconds of the signal. The average LEN energies from before collapse up to the simulated $0.5$ s after bounce are (c.f. Table 3.4 of Ref. \cite{hudepohl}): $\langle E_{\nu_e}\rangle=13$ MeV, $\langle E_{\bar{\nu}_e}\rangle=15$ MeV and $\langle E_{\nu_x}\rangle=14.6$ MeV.

Second, we also use a parametric model for neutrino emission from Pagliaroli {\it et al} \cite{pagliaroli2009}, focusing on the best-fit emission from SN1987A data and the model provides the average energies of $\langle E_{\nu_e}\rangle=9$ MeV, $\langle E_{\bar{\nu}_e}\rangle=12$ MeV 
and $\langle E_{\nu_x}\rangle=16$ MeV. This signal has a temporal structure:
\begin{equation}
F(t,\tau_1,\tau_2) = (1-e^{-{t / {\tau_1}}})e^{-{t / {\tau_2}}},
\label{eq:pagliaroli_model}
\end{equation}
where $\tau_1$ and $\tau_2$ are parameters governing the emission, representing the rise and decay timescales of the neutrino signal. The best-fit values on these parameters are $\sim 0.1$ s and $\sim 1$ s \cite{pagliaroli_ccsn} .

Here, we consider only the main interaction channel for water and scintillator detectors: the inverse beta decay (IBD) $\bar\nu_e+p \rightarrow n+e^+$. We also consider standard MSW neutrino oscillations on the flux $\Phi_{\bar{\nu}_e}$ at the detectors. The flux is an admixture of the unoscillated flavor fluxes at the source, i.e. $\Phi_{\bar{\nu}_e}=P\cdot\Phi_{\bar{\nu}_e}+(1-P)\Phi_{\bar{\nu}_x}$. Here, $P$ describes the survival probability for the $\bar{\nu}_e$. The value of $P\simeq 0$ is for the Inverted Hierarchy (IH), while $P\simeq0.7$ is for the Normal Hierarchy (NH). The expected number of IBD events can be seen in Tab.~\ref{tab:nu_models} with the reference distance of $10$~kpc.

 \begin{table}
\tbl{Average number of IBD events for a CCSN exploding at 10 kpc for the considered models detected by Super-K \protect\cite{superK}, LVD \protect\cite{lvd_det}, and KamLAND \protect\cite{kamland}, with the assumed energy thresholds ($E_\mathrm{thr}$).}
{\begin{tabular}{@{}cccccc@{}}
\toprule
Model  & Progenitor   & Super-K & LVD  & KamLAND \\
    (identifier) & Mass & ($E_\mathrm{thr}=6.5$ MeV) & ($E_\mathrm{thr}=7$ MeV) &($E_\mathrm{thr}=1$ MeV) \\\colrule
    Pagliaroli \protect\cite{pagliaroli2009} & $25\, M_\odot$ & 4120 & 224 & 255\\
     (SN1987A) &&&&\\
    \hline
     H{\"u}depohl \protect\cite{hudepohl}  & $11.2\, M_\odot$ & 2620 & 142 & 154 \\
     (Hud) &&&&\\\botrule
\end{tabular}}
\label{tab:nu_models}
\end{table}


\section{Data and Analysis Strategy\label{sec:science}}

Here, we will discuss the data and analysis for GWs, for LENs, and for a combined multimessenger search. We consider a conservative threshold on global false alarm rate (FAR) of 1/1000 years.

 
\subsection{Gravitational Wave Analysis}\label{sec:gw_analysis}

We employ the cWB\cite{Klimenko:2015ypf, Drago:2020kic}\footnote{cWB home page, \url{https://gwburst.gitlab.io/}; \\ public repositories, \url{https://gitlab.com/gwburst/public}\\ documentation, \url{https://gwburst.gitlab.io/documentation/latest/html/index.html}.} algorithm for the GW data analysis, a widely used software in the LIGO-Virgo-KAGRA collaborations; which is also used for the study of CCSN targeted-search \cite{SNTargeted2016,em_gw_ccsne}. The cWB software does not need any waveform templates; it combines coherently the excess energy of the data from the involved GW detectors. A maximum likelihood analysis is used to search for the GW candidates and their parameters. The significance is estimated comparing foreground the detection statistics $\rho$ (of the candidates) with a background distribution from the time-shift procedure. We simulate Gaussian noise with a spectral sensitivity based on the expected Advanced LIGO \cite{Abbott:2020qfu} and Advanced Virgo detectors \cite{TheLIGOScientific:2014jea, TheVirgo:2014hva}. We simulate $\sim16$ days of data and $\sim20$ years of background livetime.

Waveforms from the models in Sec.~\ref{sec:GWemission} have been simulated for several distances: 5, 15, 20, 50, 60, 700 kpc. In terms of sky direction, for the shorter distances (5, 15, 20) we use the Galactic model given in \cite{marektesi}, whereas for the longer (50 and 60 kpc) distances we take fixed directions, either towards the Large or Small Magellanic Clouds, and additionally for the 700 kpc we use the Andromeda direction. We focus mainly on the 5-60 kpc distances for the multimessenger analysis, while the distance between 60 and 700 kpc are used for the efficiency curve\footnote{For the simulations between 60 and 700 kpc, we use the Andromeda position, though no known astronomical objects are expected around that area.}. 
We inject 1 GW signal per 100 seconds, wide enough for two consecutive waveforms. In terms of sky positions, we apply $\sim 2500$ different realizations over all the sky direction for each distance and model . 

\begin{figure}[!ht]
    \centering
    \includegraphics[width=.7\linewidth]{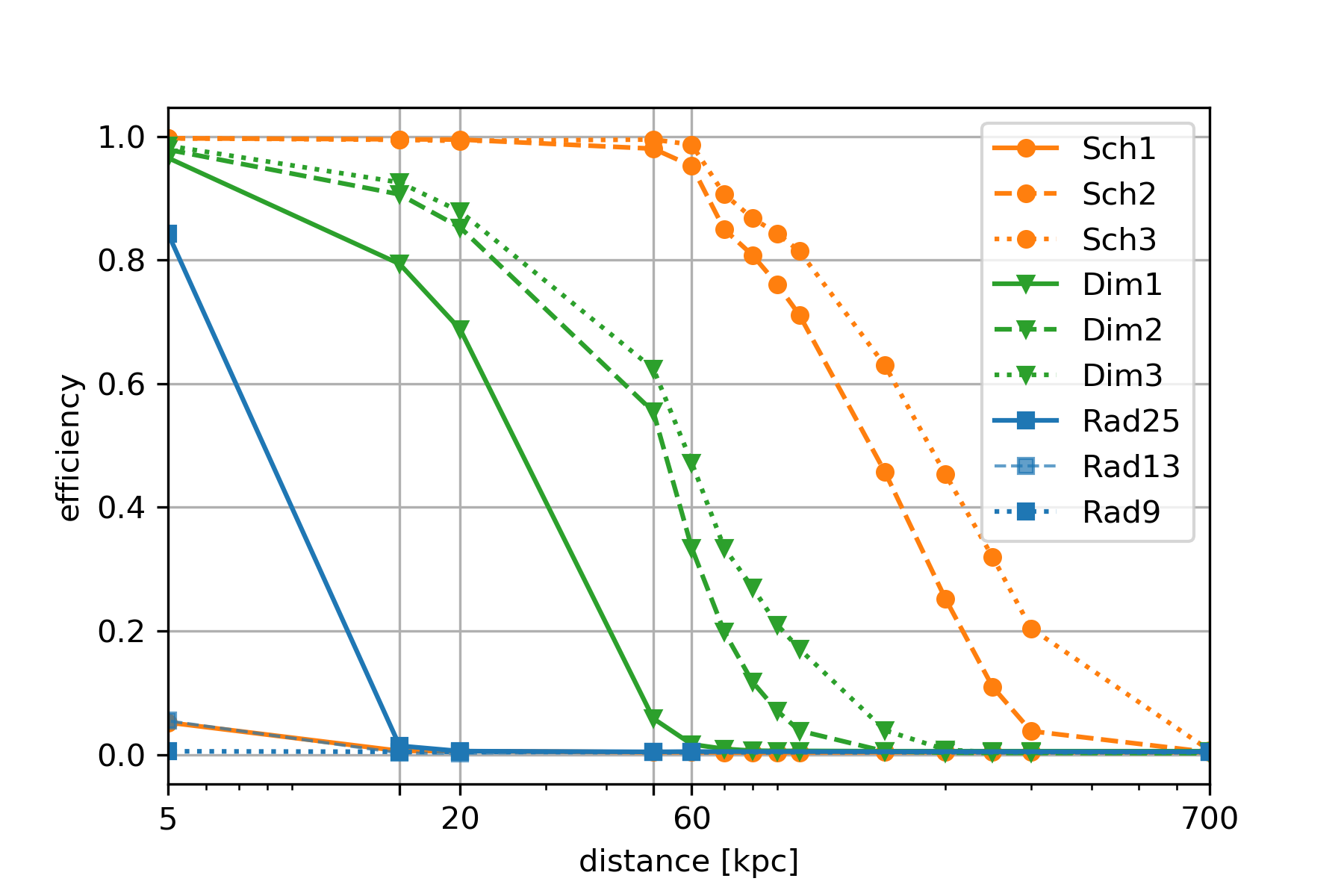} 
    \label{fig:cWBeff1}
      \caption{Efficiency curve of GW search from Advanced LIGO and Advanced Virgo detectors for the various models in Tab. \protect\ref{tab:gw_models}, given a FAR threshold of $864$/day. Sch1-3 are for the Scheidegger model with different frequency peaks, Dim1-3 are for the Dimmelmeier model with different $h_\mathrm{rss}$ values, and Rad1-3 are for the Radice model with different progenitor masses}
      \label{fig:gw_1987_60}
    \end{figure}

We choose a threshold of $864$ per day for GW candidates to be used for our multimessenger analysis, to satisfy the requirement of the combined FAR of 1/1000 years. The efficiency curves for each distance can be seen in Fig.~\ref{fig:gw_1987_60} as the ratio of the number of recovered injections ($\mathrm{FAR_{GW}}<864/$ day) to the $\sim 2500$ total injections.  

\subsection{Neutrino Analysis: Expanding the Neutrino Detection Horizon}
\label{subsec:neutrino}

The standard LEN analysis for CCSNe \cite{Agafonova2014,Ikeda2007,Abe2016} employs a binning in a time series data set with a sliding time window of $w=20$ seconds. The group of events in a bin is called a \textit{cluster} and the number of events in a cluster is called multiplicity $m$. We can assume the multiplicity distribution of background events follows a Poisson distribution. For each $i$-th cluster, we calculate the \textit{imitation frequency} ($f^\mathrm{im}$), which is correlated with the significance, defined as,
\begin{equation}
f^\mathrm{im}_i (m_i)=N\times \sum_{k=m_i}^\infty P(k),
\label{eq:fim_prob}
\end{equation}
where $P(k)$ is the Poisson term: the probability in which a cluster of multiplicity $k$ is simply due to the background,          
\begin{equation}
P(k)=\frac{(f_\mathrm{bkg}w)^k e^{-f_\mathrm{bkg}w}}{k!},
\label{eq:poiss_pdf}
\end{equation}
with $N=8640$ is the total number of bins in one day, calculated from the $10$-s overlapping window between two consecutive bins, to tackle the boundary problems. This $f^{im}$ is basically called FAR in the GW community.


From our previous work \cite{Casentini2018}, we can exploit the temporal behavior of LEN signals with an additional cluster parameter called $\xi_i\equiv \frac{m_i}{\Delta t_i}$, with $\Delta t_i$ as the $i$-th cluster duration: the time difference between the last and the first event. Clearly, $\Delta t_i$ has a maximum value of the bin width itself, which is 20 seconds. A cluster is considered when $m_i\geq2$, thus $\xi_i\geq0.1$.

\begin{figure}[!ht]
\centering
\includegraphics[width=.7\linewidth]{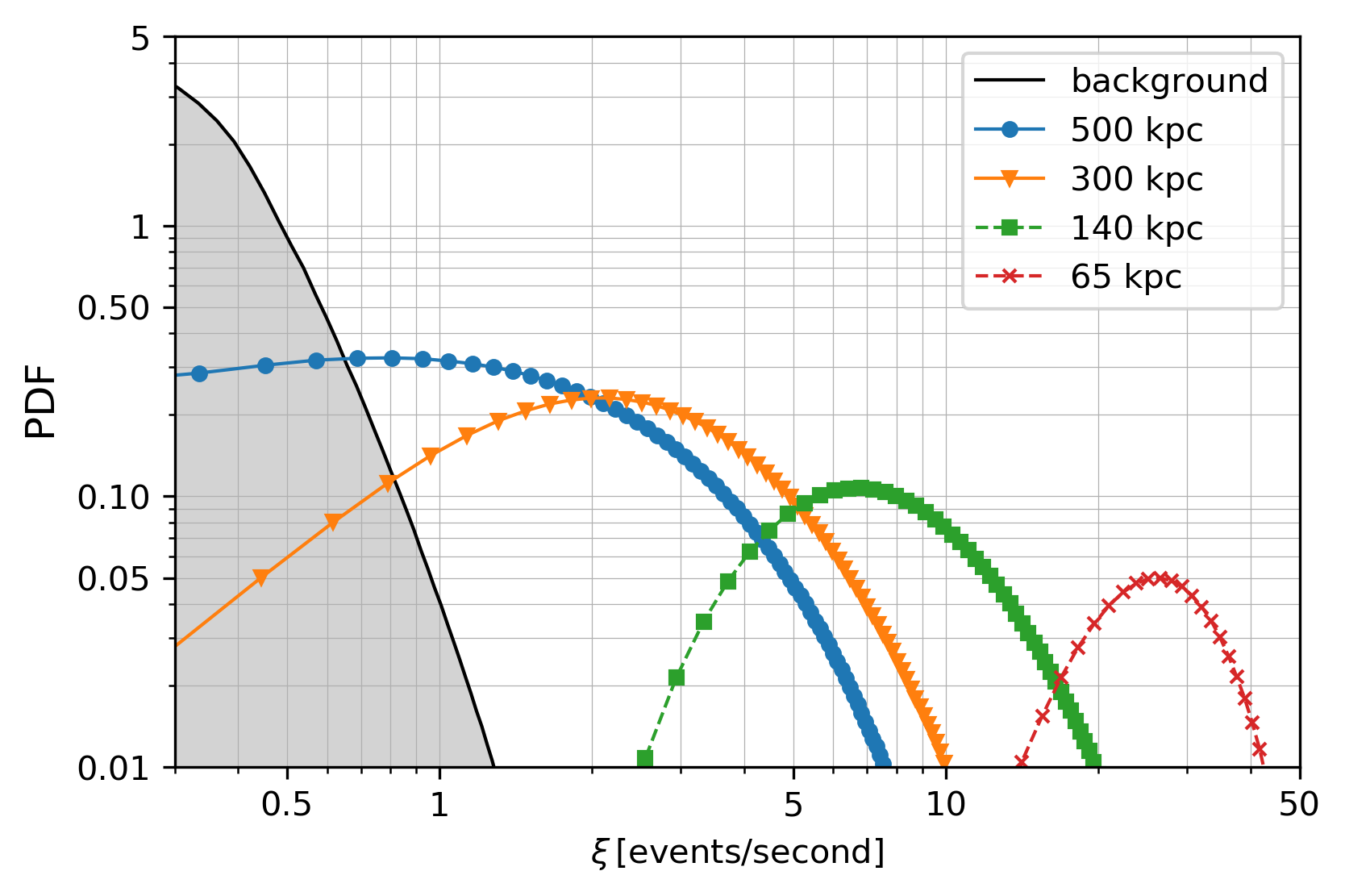}
\caption{PDF of $\xi$ values for background (black line) and signal clusters (colored-lines) for different distances in the case of the Super-K detector. Data are from \protect\cite{Casentini2018}.}
\label{fig:pdf}
\end{figure}

Previous work \cite{Casentini2018} uses $\xi$ as a cut value. In this work we develop further in order improve the estimation of the imitation frequency. Particularly, it is formulated a new modified \textbf{2-parameter} ($m$ and $\xi$) imitation frequency, called $F^\mathrm{im}$:
\begin{equation}
F^\mathrm{im}_i (m_i,\xi_i)=  N \times  \sum_{k=m_i}^\infty  P(k,\xi_i),
\label{eq:newfim_0}
\end{equation}
where the term $P(k,\xi_i)$ is the joint probability of a cluster with multiplicity $k$ \textbf{and} $\xi_i$ substituting $P(k)$ previously (Eq.~\ref{eq:poiss_pdf}). Finally, it is convenient to write (see c.f. App. A of \cite{Halim2021} and c.f. Sec.~7.1. of \cite{halimtesi} for more detail),
\begin{equation}
F^\mathrm{im}_i(m_i,\xi_i)= N \times \sum_{k=m_i}^\infty P(k) \int_{\xi=\xi_i}^\infty \mathrm{PDF}(\xi\geq\xi_i|k) d\xi.
\label{eq:newfim_1}
\end{equation}
Fig.~\ref{fig:pdf} shows the $\xi$ probability density function (PDF) for the Super-K detector (as an example). The black solid line represents the PDF of $\xi$ values due to the background, i.e. $\mathrm{PDF}(\xi | k)$.


%
%
%
%

We simulate $\sim10$ years of background data for each neutrino detector with the assumption that the background frequencies $f_\mathrm{bkg}$: $0.012$ Hz for Super-K \cite{Abe2016}, $0.015$ Hz for KamLAND \cite{kam_bg}, and $0.028$ Hz for LVD \cite{Agafonova2014}. Moreover, the CCSN simulated signals (see Sec.~\ref{sec:sources}) are injected into the background data for each detector model (the starting times are simultaneous with the GW injections, with reasonable time delay due to detectors' positions) and for each source distance. These clusters are extracted via the Monte Carlo method and the injection rate is 1 per day.

\begin{figure}[!ht]
    \centering
      \includegraphics[width=.7\linewidth]{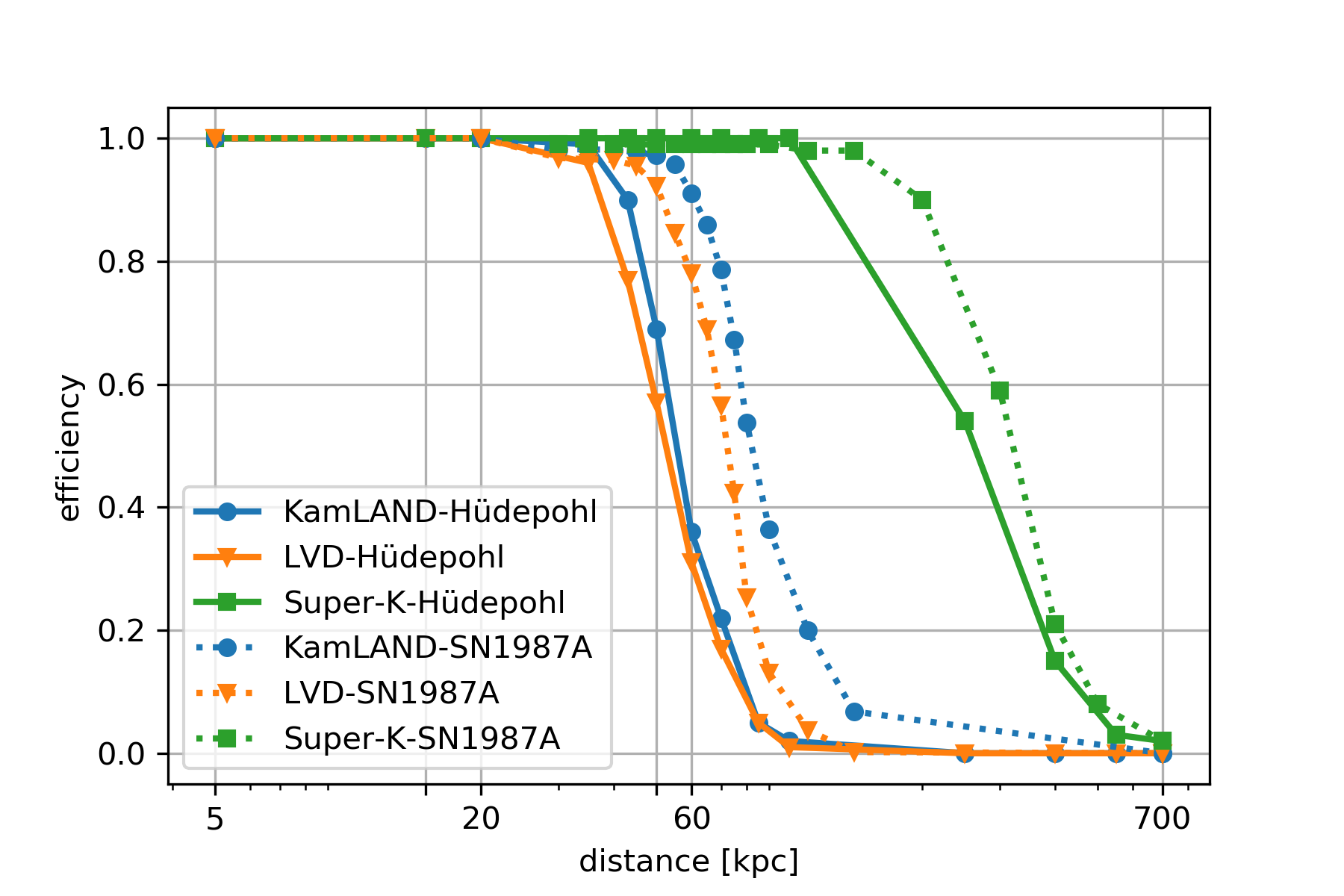}
      \caption{The efficiency curves of LEN detectors for the Hud (continuous lines) and SN1987A (dashed lines). The imitation frequency threshold is 1/day.}
      \label{fig:Efficiency}
    \end{figure}

The clusters are considered signal candidates if their standard $f^\mathrm{im}\leq 1/$day, which has been set in order to reach the global FAR of $1/1000$ years. The efficiency for each distance $D$ can be defined as,
\begin{equation}
\eta(D) = \frac{N_\mathrm{r,s}(D)}{N_\mathrm{inj,s}(D)}.
\label{eq:det_eff}
\end{equation}
where $N_\mathrm{r,s}$ is the number of candidates and $N_\mathrm{inj,s}$ is the total number of injections. With the recipes up to now, the efficiency curves can be constructed (see Fig.~\ref{fig:Efficiency}).


\subsection{Multimessenger Analysis}
\label{FAR}

The ultimate goal of our work is to construct a multimessenger analysis, combining both LEN and GW data sets (green boxes in Fig.~\ref{fig:GWnu_scheme}). We simply perform a temporal-coincidence analysis between the GW and LEN lists and the statistical significance can be estimated by the combination of its FAR. The joint coincidences are defined as ``CCSN candidates''.

The $\mathrm{FAR_{GW}}$ is acquired via the time-shifting method (Sec.~\ref{sec:gw_analysis}) while, the $\mathrm{FAR_{\nu}}$ from LEN is obtained by the product method based on SNEWS \cite{Antonioli2004} i.e.,
\begin{equation}
\mathrm{FAR_{\nu}}=\mathrm{Nd}\times w_{\nu}^{\mathrm{Nd}-1}\prod_{i=1}^{\mathrm{Nd}}F^\mathrm{im}_i,
    \label{eq:farLEN}
\end{equation}
where $\mathrm{Nd}$ is the number of LEN detectors, $w_{\nu}$ is the time window of coincidence analysis, and $F^\mathrm{im}_i$ is the 2-parameter imitation frequency of the clusters.

All in all, the multimessenger {$\mathrm{FAR_{glob}}$} from ``CCSN candidates'' can be written\footnote{Thorough discussion on the choice of coincidence analysis can be seen in \cite{halimtesi}.},
\begin{equation}
    \mathrm{FAR_{glob}}=\mathrm{Net}\times w_c^{\mathrm{Net}-1}\prod_{X=1}^\mathrm{Net}\mathrm{FAR}_X,
    \label{eq:jointfar}
\end{equation}
where $\mathrm{Net}$ is the number of sub-networks, $w_c$ is the temporal coincidence window between GW and LEN, and $\mathrm{FAR}_X$ is the \textit{false-alarm-rate} ($X=\{\mathrm{\nu,\,GW}\}$). Thus, the \textit{false-alarm-probability} taking into account Poisson statistics can be written,
\begin{equation}
    \mathrm{FAP}=1-e^{-\mathrm{FAR}\times\mathrm{livetime}},
    \label{eq:jointfap}
\end{equation}
where ``$\mathrm{livetime}$'' is the common observing time among the network.

Eq.~\ref{eq:jointfar} and \ref{eq:jointfap} are used to compare the performance of our 2-parameter method (Eq. \ref{eq:newfim_1}) with the standard 1-parameter method (Eq. \ref{eq:fim_prob}). This performance is discussed as efficiency values, analogous with Eq.~\ref{eq:det_eff}. For the multimessenger step, we define a ``detection'' in a network when $\mathrm{FAP}\geq 5\sigma$\footnote{$5\sigma\approx 5.7\times 10^{-7}$}.






\section{Results \label{sec:results}}
Here, we discuss the results of our method. First, we mention the single-detector neutrino analysis performance. Next, we show the sub-network of neutrino detectors. Finally, we provide the global network of GW-LEN multimessenger analysis.


\subsection{Improvement of the single-detector LEN analysis}\label{sec:single_det}
 
We work our method on the simulated single-detector data for KamLAND, LVD, and Super-K. To show the improvement, we discuss in the following, as a leading example, the case of the KamLAND detector. We consider a CCSN at 60 kpc with the LEN profile following SN1987A model (the first row of Tab.~\ref{tab:nu_models}). We perform a {10 years} of KamLAND simulated background data and we inject randomly the simulated LENs with the rate of 1/day, thus, 3650 total injections.

\begin{figure}[!ht]
    \centering
      \includegraphics[width=.7\linewidth]{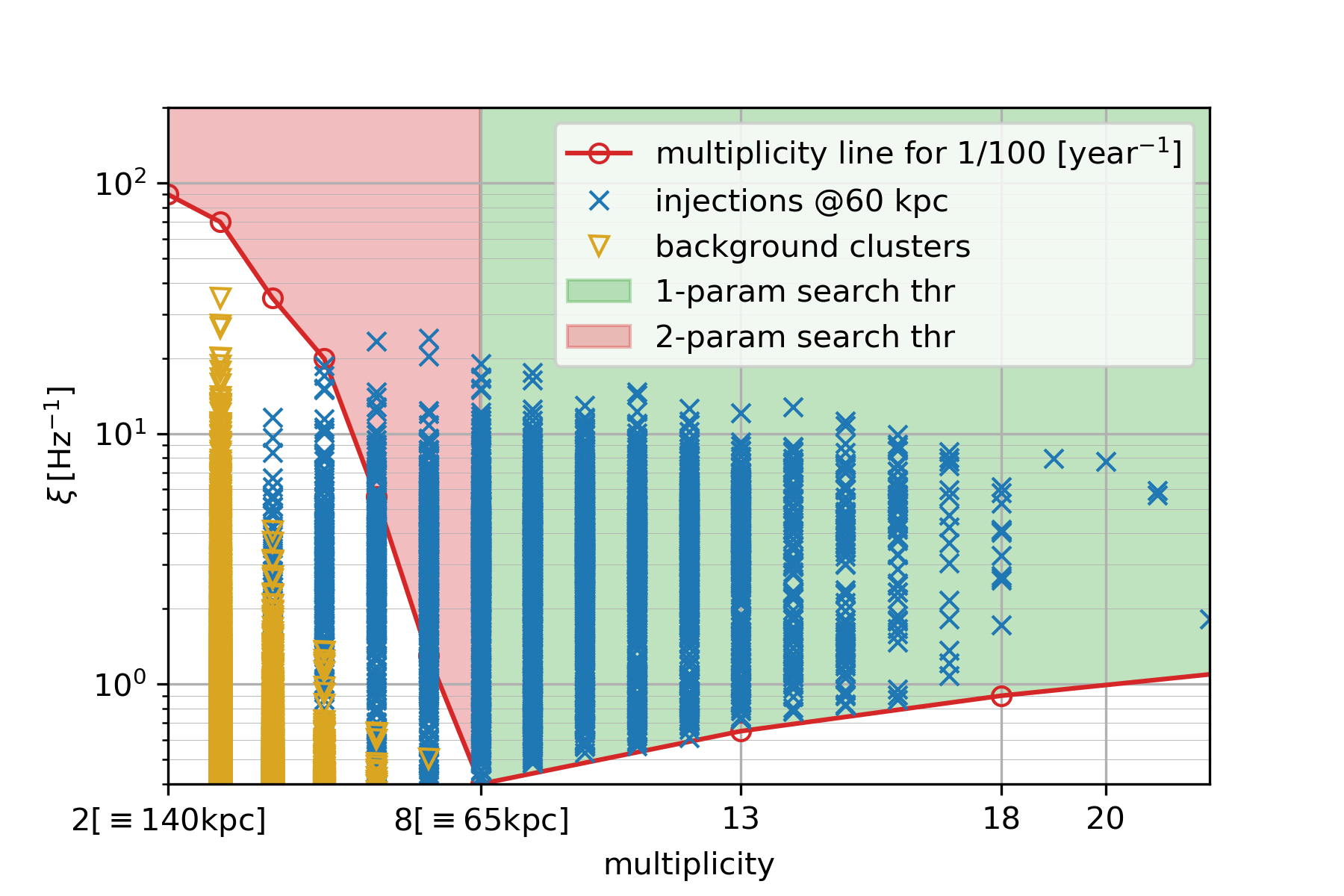}
      \caption{The $\xi$-multiplicity map for KamLAND with the simulated background (yellow-triangle) and injection (blue-cross) clusters generated with the SN1987A model at 60 kpc.}
      \label{fig:kamland_60kpc}
    \end{figure}

All clusters, either due to background or injections, are plotted in Fig. \ref{fig:kamland_60kpc} in a $\xi$ {vs} multiplicity plane. Blue crosses represents the clusters due to injections while the yellow inverted-triangles are due to background. Because of the statistical Poisson fluctuation in the Monte Carlo simulation method, the injection clusters may have various multiplicity values. Besides, background events are also fluctuating in this 20-sec window.

We estimate the associated imitation frequency (or $\mathrm{FAR}_\nu$\footnote{As previously stated in Sec. \ref{subsec:neutrino}, the imitation frequency is actually the $\mathrm{FAR}$. So, $\mathrm{FAR}_\nu$ is imitation frequency for KamLAND data, either $f^\mathrm{im}$ or $F^\mathrm{im}$ depending on the context.}) for each recovered cluster. This imitation frequency in 1-parameter analysis is a function of only the cluster multiplicity via Poisson statistics. To satisfy the SNEWS requirement of $\mathrm{FAR_\nu}\leq1/100$ year, the multiplicity threshold is 8. This means clusters should lie on the green area of Fig.\ref{fig:kamland_60kpc} to satisfy the standard 1-parameter requirement. This limit on the multiplicity could be understood as a KamLAND horizon for CCSN search of $\simeq 65$ kpc with the model based on SN1987A, in fact it is the average multiplicity for this CCSN configuration: $\langle m_i\rangle=8$.

On the other hand, the imitation frequency of our 2-parameter method requires each cluster to satisfy the requirement of both multiplicity and the $\xi$ value following Eq.~\ref{eq:newfim_0}. That equation gives us the red line in Fig.~\ref{fig:kamland_60kpc} for $\mathrm{FAR_{\nu}}=1/100$ years\footnote{The threshold corresponding to the current SNEWS requirement.} required for the 2-parameter method. All clusters above the red line pass the 2-parameter method requirement. This also relax the requirement allowing multiplicity lower than 8 as long as a specific $\xi$ value is satisfied. Thus, the red area in Fig.~\ref{fig:kamland_60kpc} represents the improvement area of our 2-parameter method comparing with the 1-parameter. In addition, it is clear from the figure that all simulated background clusters (yellow triangles) are well below the red threshold line.

Quantitatively, Fig.~\ref{fig:kamland_60kpc} shows an increase of the detection efficiency with details given in the first row of Tab.~\ref{tab:kam_lvd_eff}. The KamLAND efficiency at 60 kpc is improved from $73\%$ by the standard 1-parameter method to $83\%$ by our 2-parameter method. Moreover, we mention also that there are 75198 noise triggers in this data set and none of them have FAR lower than 1/100 years for both methods. The efficiency increase could also mean we can expand the detection horizon of the detector since the expected multiplicity is proportional to the inverse of squared-distance\footnote{The number of events is basically a flux.}. The results for KamLAND are representative for all the scenarios we investigated. Similar improvements can be seen for other scenarios (see c.f. App.~C of \cite{Halim2021}). 



\subsection{The LEN detector sub-network}
We apply our method for the sub-network of LEN detectors: KamLAND and LVD, since their efficiency curves are compatible; see Fig.~\ref{fig:Efficiency}. The threshold for LEN detector network is $5\sigma$ in $\mathrm{FAP}_\nu$ (Eq. \ref{eq:jointfap}). 
We compare the efficiencies in the second and third row of Tab.~\ref{tab:kam_lvd_eff}.





\begin{table}
\tbl{Efficiency ($\eta$) comparison between 1-parameter and 2-parameter method for analysis of KamLAND (with the SN1987A model) and KamLAND-LVD (with the Hud model) and for $\mathrm{FAP_\nu}>5\sigma$.}
{\begin{tabular}{@{}ccccc@{}}
\toprule
Analysis & Model & Distance $\mathrm{[kpc]}$ &  $\eta_\mathrm{1param}$ & $\eta_\mathrm{2param}$ \\
&     & $\left[>5 \sigma \right]$ &  $\left[> 5 \sigma\right]$ \\\colrule
     
 KamLAND & SN1987A & 60 &  \cellcolor{magenta!30}\textbf{2665/3654=72.9\%}  & \cellcolor{yellow!70}\textbf{3026/3654=82.8\%}   \\\colrule

KamLAND-LVD & Hud&50 & \cellcolor{magenta!30}\textbf{47/108=43.5\%} & \cellcolor{yellow!70}\textbf{59/108=54.6\%}\\
KamLAND-LVD &Hud &     60 &  \cellcolor{magenta!30}\textbf{19/107=17.8\%} & \cellcolor{yellow!70}\textbf{28/107=26.2\%}\\\botrule
\end{tabular}}
\label{tab:kam_lvd_eff}
\end{table}

%

At a distance of 50 kpc with a $\mathrm{FAR_\nu}\le 1/100$ years, we can detect $12\%$ and $26\%$ CCSNe for LVD and KamLAND, respectively. Meanwhile, if the detectors are in a network looking for coincidences within $w_\nu$, it is possible to recover $\sim 43\%$ with $5\sigma$ threshold. When we start using 2-parameter method taking into account the $\xi$ distribution for each detector, this efficiency grows to $\sim 55\%$.

Analogously, for CCSNe at 60 kpc, the signals with $\mathrm{FAR_\nu}\le 1/100$ years can be detected only $3\%$ and $7\%$ for LVD and KamLAND. Whereas, as a network this efficiency increases to $18\%$ and to $26\%$ for 1-parameter and 2-parameter method, respectively fo $5\sigma$ threshold.


In order to quantify the result, we will mention the results for the SN1987A model with a CCSN at $60$ kpc recovered by the LVD-KamLAND network. The improvement in efficiency is from {$85\%$ to $93\%$}. These results can be compared with first row of Tab. \ref{tab:kam_lvd_eff} for the single detector analysis.



\subsection{The multimessenger network}

In the multimessenger network, the temporal coincidence window is $w_c=10$ seconds between GW and LEN lists. We use the $5\sigma$ threshold to claim a multimessenger ``detection'' of injected signals. We focus on emphasizing the improvement of combining GWs and LENs, which means when the detection efficiency of both LEN and GW detectors is less than $100\%$. In other words, it is to combine detectors with comparable detection efficiencies. The horizon of the GW network is highly influenced by the assumed GW emission model, see Fig. \ref{fig:gw_1987_60}. In this case, the Dim2 model (see Tab. \ref{tab:gw_models}) has the GW detection horizon comparable with the LVD and KamLAND detectors, which is more or less the Large Magellanic Cloud distance. Thus, we perform the method for the global network of LIGO-Virgo, LVD, and KamLAND. Results for other GW models can be seen in c.f. App.~B of \cite{Halim2021}.


\begin{figure}[!ht]
    \centering
      \includegraphics[width=.7\linewidth]{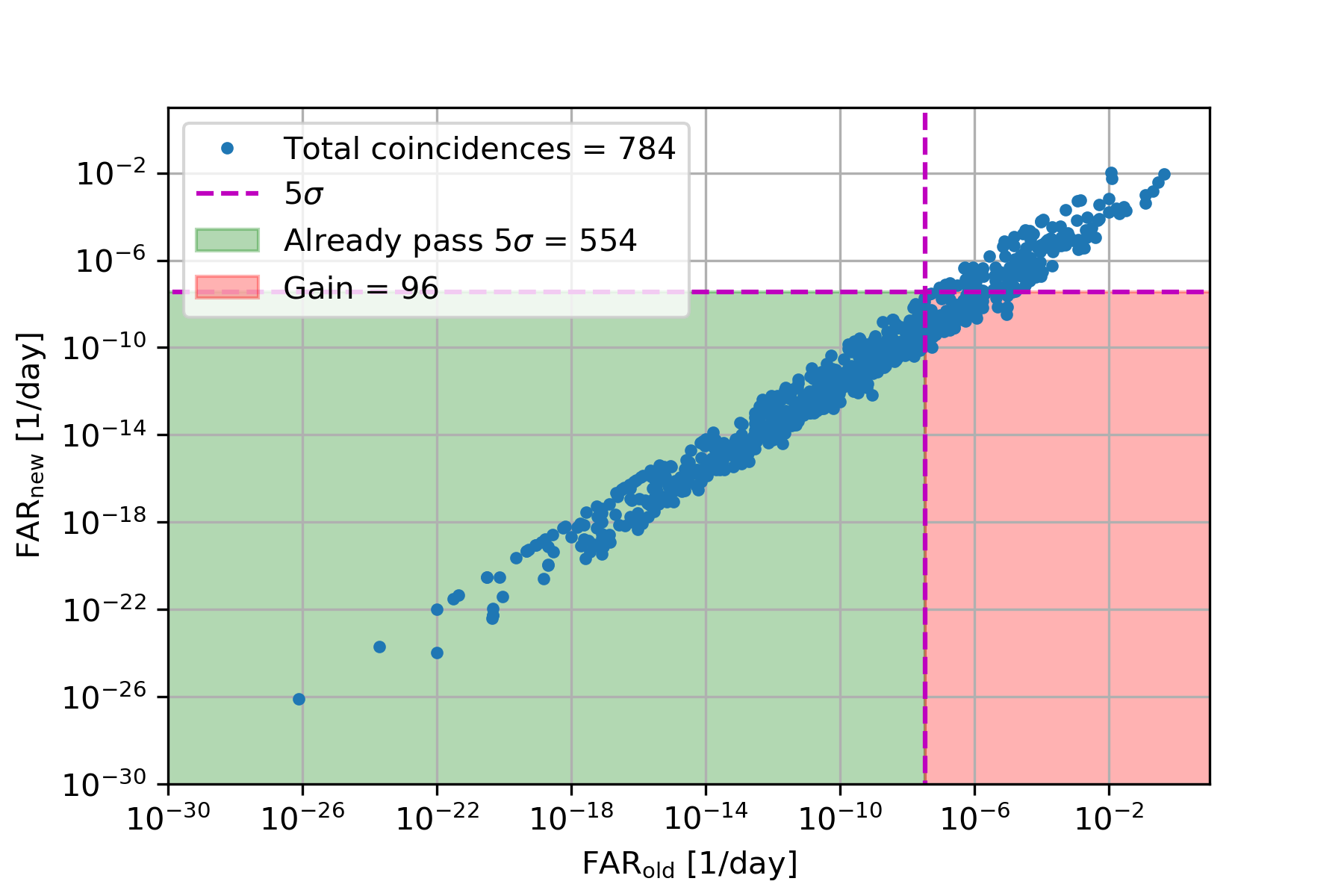}
      \caption{{The joint FAR of GW-LEN candidates obtained with the 2-parameter method ($\mathrm{FAR_{new}}$ $y$-axis) vs the 1-parameter ($\mathrm{FAR_{old}}$ $x$-axis) with the \textbf{KamLAND} (SN1987A-model) and HLV (Dim2-model) at 60 kpc.}}
      \label{fig:gw_1987_60_dou}
    \end{figure}%
    
For the coincidence analysis of KamLAND detector with the HLV (Hanford, Livingston, Virgo) GW network, we can see the FAR comparison in Fig.~\ref{fig:gw_1987_60_dou} for the case of CCSNe at $60$ kpc and with the LENs from SN1987A model and the GW from Dim2 model. The magenta dashed-line is $5\sigma$ significance threshold. Either $\mathrm{FAR_{old}}$ or $\mathrm{FAR_{new}}$ values lower than this line mean they pass $5\sigma$. Thus, all clusters (blue dots) lie on the green area are recovered by the 1-parameter method. While, the red area is the improvement: the clusters recovered by the 2-parameter but not by the 1-parameter. The first row of Tab.~\ref{tab:dimkamlvd} gives us the efficiency of the 1-parameter and the 2-parameter method. This yields an additional {$\sim12\%$} of signals from the 2-parameter that are discarded by the 1-parameter. 


     \begin{figure}[!ht]
    \centering
    \includegraphics[width=.7\linewidth]{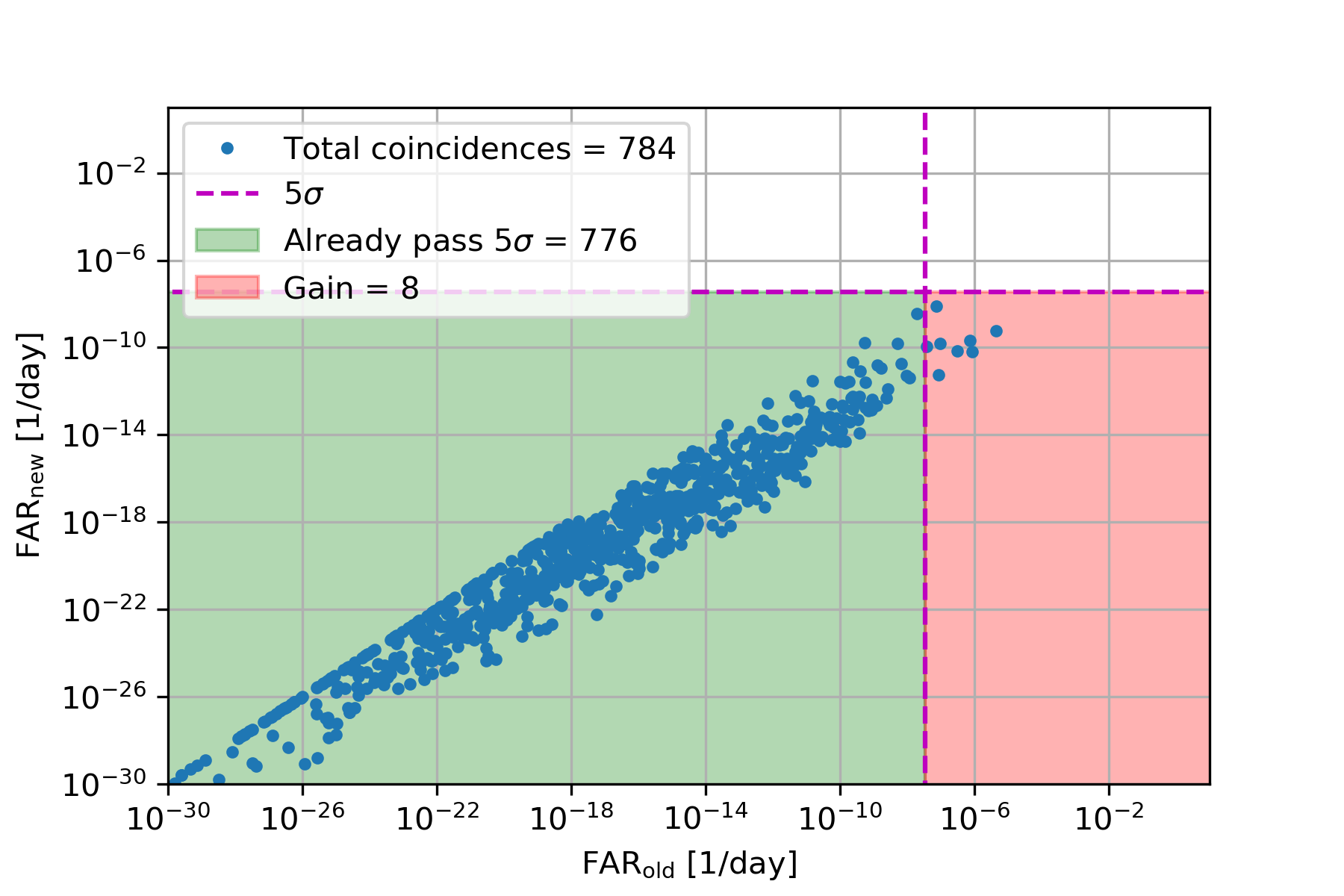}
      \caption{The joint FAR of GW-LEN candidates obtained with the 2-parameter method ($\mathrm{FAR_{new}}$ $y$-axis) vs the 1-parameter ($\mathrm{FAR_{old}}$ $x$-axis) with the \textbf{KamLAND-LVD} (SN1987A-model) and HLV (Dim2-model) at 60 kpc.}
      \label{fig:gw_1987_60_tri}
    \end{figure}%

    \begin{table}
\tbl{Efficiency comparison of the 1-parameter ($\eta_\mathrm{1param}$) and the 2-parameter ($\eta_\mathrm{2param}$) method for Fig.~\protect\ref{fig:gw_1987_60_dou} (upper row) and \protect\ref{fig:gw_1987_60_tri} (lower row). The first column indicates the specific network of detectors and models. The second column shows the GW results with the threshold on the $\mathrm{FAR_{GW}}\mathrm{<864/day}$). The third and last columns are the efficiency with $>5\sigma$ significance.}
{\begin{tabular}{@{}cccc@{}}
\toprule
Network $\&$ Type  & Recovered   & $\eta_\mathrm{1param}$ & $\eta_\mathrm{2param}$  \\
    of Injections & $\mathrm{FAR_{GW}}<864/\mathrm{d}$ &  $\left[>5\sigma\right]$ & $\left[>5\sigma\right]$  \\\colrule
HLV-KAM  & 784/2346= &  554/784= &  650/784=  \\
    (Dim2-SN1987A) & 33.4\% & \cellcolor{magenta!30}\textbf{70.7\%} &  \cellcolor{yellow!70}\textbf{82.9\%} \\
    \hline
    HLV-KAM-LVD  & 784/2346= & 776/784= &  784/784=   \\
    (Dim2-SN1987A) & 33.4\% & \cellcolor{magenta!30}\textbf{99.0\%}  &  \cellcolor{yellow!70}\textbf{100\%}  \\\botrule
\end{tabular}}
\label{tab:dimkamlvd}
\end{table}

The cWB performs and analyses 2346 GW injections, and out of that, 784 of them have $\mathrm{FAR_{GW}<864/day}$. These candidates' significances are far too low to be even taken as \textit{sub-threshold} triggers. Then, temporal coincidence analysis is performed with the input of these GW triggers and the list of LEN clusters. There are {554} coincident candidates that pass $5\sigma$ ($\sim 71\%$ of the GW triggers) with the standard 1-parameter method. However, if we employ the new 2-parameter method, we can get additional 110 recovered signals. This means the efficiency increases to $\sim 83\%$ (the first row of Tab.~\ref{tab:dimkamlvd}).

We also study further this method involving the LVD detector to be in the triple-detector configuration. We show the result in Fig.~\ref{fig:gw_1987_60_tri} and quantitative summary in the second row of Tab.~\ref{tab:dimkamlvd}. The improvement of our method seems less evident. This is clear due to the fact that the efficiency cannot go beyond the maximum value of $33.4\%$, where all the GW triggers from cWB are coincident with the list of LEN clusters with $>5\sigma$ significance.

\begin{figure}[!ht]
    \centering
      \includegraphics[width=.7\linewidth]{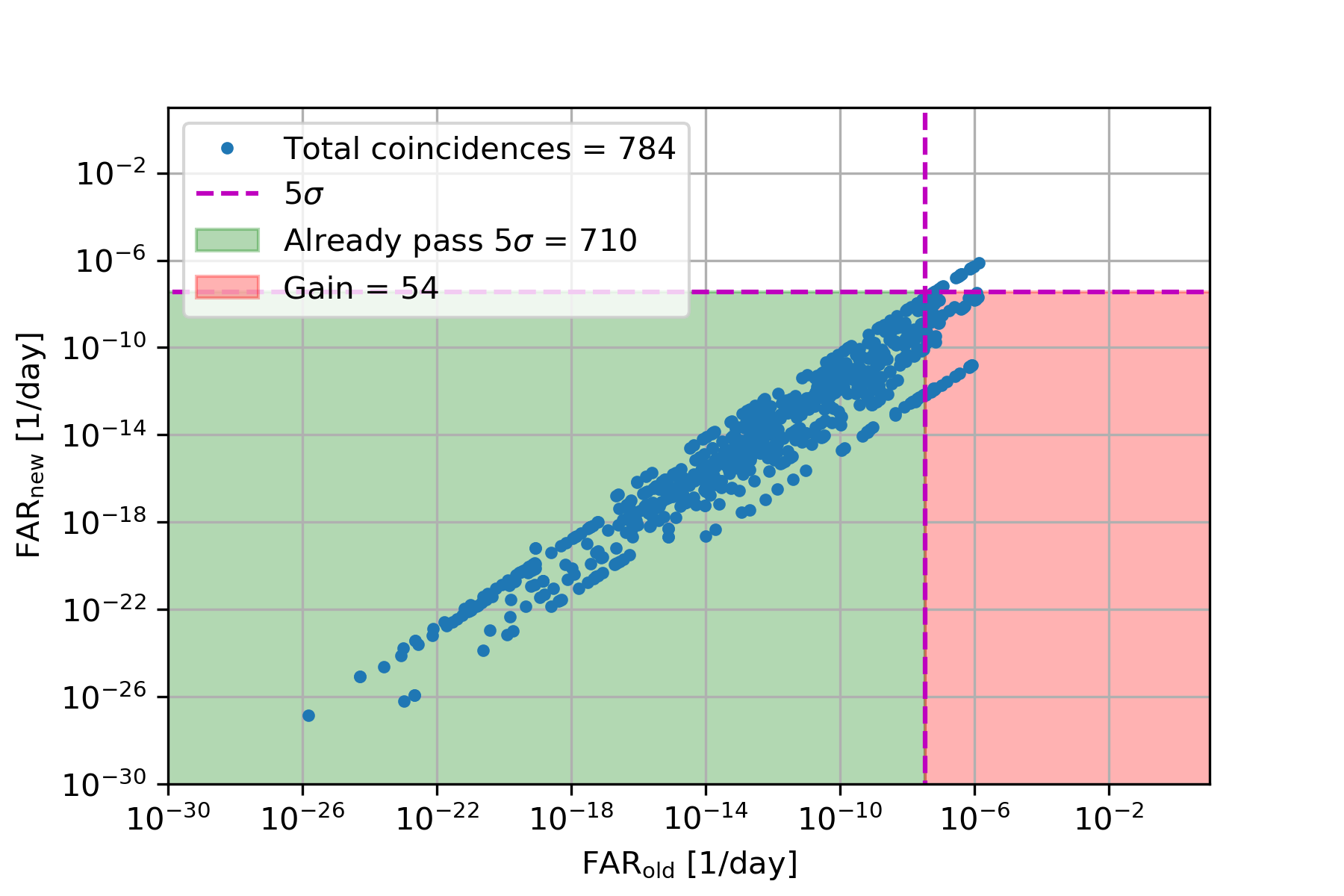}
      \caption{The joint FAR of GW-LEN candidates obtained with the 2-parameter method ($\mathrm{FAR_{new}}$ $y$-axis) vs the 1-parameter ($\mathrm{FAR_{old}}$ $x$-axis) with the {KamLAND-LVD} (\textbf{Hud-model}) and HLV (Dim2-model) at 60 kpc.}
      \label{fig:dim2_no_60}
    \end{figure}

We then provide the result by using the Hud model in exchange of the SN1987A model. The FARs can be seen in Fig.~\ref{fig:dim2_no_60} and the efficiency values in Tab.~\ref{tab:dimkamlvd_hud}. This 2-parameter method gives us $\sim7\%$ more recovered signals. We found that on average, the FARs the injections are $O(10^3)$ smaller with the 2-parameter than the 1-parameter for both emission models.

    \begin{table}
\tbl{Efficiency comparison of the 1-parameter ($\eta_\mathrm{1param}$) and the 2-parameter ($\eta_\mathrm{2param}$) method for Fig.~\protect\ref{fig:dim2_no_60}. The columns are analogous to Table \protect\ref{tab:dimkamlvd}.}
{\begin{tabular}{@{}cccc@{}}
\toprule
Network $\&$ Type & Recovered   & $\eta_\mathrm{1param}$ & $\eta_\mathrm{2param}$  \\
    of Injections & $\mathrm{FAR_{GW}}<864/\mathrm{d}$ &  $\left[>5\sigma\right]$ & $\left[>5\sigma\right]$  \\\colrule
    HLV-KAM-LVD & 784/2346=  & 710/784=  & 764/784=    \\
    (Dim2-Hud) & 33.4\%  & \cellcolor{magenta!30}\textbf{90.6\%} & \cellcolor{yellow!70}\textbf{97.5\%}  \\\botrule
\end{tabular}}
\label{tab:dimkamlvd_hud}
\end{table}

All in all, let us summarize the results. Fig.~\ref{fig:gw_1987_60} for Dim2 model shows that the GW network HLV, by applying a threshold $\mathrm{FAR_{GW}}\le 864$/day, recovers about {$\sim33\%$} of the injected signals at 60 kpc. Notice that these GW triggers are far from statistically significant and clearly $0\%$ passes a $5\sigma$ threshold. With this global network analysis and with our 2-parameter method, the multimessenger detection efficiency grows to $\sim 33\%$. If taking into account the weaker emission like Hud model, the detection efficiency arrives at the value of {$33.4\%\cdot97.5\%=32.6\%$}.

%


\section{Conclusion \label{sec:conclusion}}


This paper is a timely description of the multimessenger strategy to combine GWs and {LENs} to search for CCSNe. More detail results will be discussed in a forthcoming paper \cite{Halim2021} that has been accepted by JCAP.

Various models of GW and LEN emissions have been considered and used for a set of simultaneous injections into the background from the considered detectors. The data are then analysed with our proposed 2-parameter method comparing with the standard 1-parameter method. The performance is exercised at the level of efficiency estimation between these two methods for various analysis steps. In general a multimessenger analysis can give a better efficiency to low-significant signals.


We highlight the improvement of the neutrino analysis with the $\xi$ parameter in terms of {FAR} and FAP. This method could be of interest of the online alert system like SNEWS2.0.





{All in all, the multimessenger campaign between GWs and LENs with the proposed method can increase the overall efficiency. Due to the higher sensitivity in LEN detectors, we can also do a targeted search in the (hopefully near) future to search for GWs from CCSNe.}

\section{Acknowledgement}
The authors gratefully acknowledge the support of the NSF for the provision of computational resources. The work of GP is partially supported by the research grant number 2017W4HA7S ``NAT-NET: Neutrino and Astroparticle Theory Network'' under the program PRIN 2017 funded by the Italian Ministero dell'Istruzione, dell'Universit\`{a} e della Ricerca (MIUR).

\bibliographystyle{ws-procs961x669}
\bibliography{ws-pro-sample}

\begin{thebibliography}{10}

\bibitem{Halim2021}
O.~Halim, C.~Casentini, M.~Drago, V.~Fafone, K.~Scholberg, C.~F. Vigorito and
  G.~Pagliaroli, Multimessenger analysis strategy for core-collapse supernova
  search: Gravitational waves and low-energy neutrinos  (2021),
  \href{https://arxiv.org/abs/2107.02050}{arXiv:2107.02050}.

\bibitem{pagliaroli_PRL}
G.~Pagliaroli, F.~Vissani, E.~Coccia and W.~Fulgione, Neutrinos from supernovae
  as a trigger for gravitational wave search, {\em Phys. Rev. Lett.} {\bf 103},
  p. 031102 (Jul 2009).

\bibitem{leonor}
I.~Leonor, L.~Cadonati, E.~Coccia, S.~D'Antonio, A.~D. Credico, V.~Fafone,
  R.~Frey, W.~Fulgione, E.~Katsavounidis, C.~D. Ott, G.~Pagliaroli,
  K.~Scholberg, E.~Thrane and F.~Vissani, Searching for prompt signatures of
  nearby core-collapse supernovae by a joint analysis of neutrino and
  gravitational wave data, {\em Classical and Quantum Gravity} {\bf 27}, p.
  084019 (apr 2010).

\bibitem{hirata1987}
K.~Hirata {\em et~al.}, Observation of a neutrino burst from the supernova
  sn1987a, {\em Phys. Rev. Lett.} {\bf 58}, 1490 (Apr 1987).

\bibitem{Bionta1987}
R.~M. Bionta {\em et~al.}, Observation of a neutrino burst in coincidence with
  supernova 1987a in the large magellanic cloud, {\em Phys. Rev. Lett.} {\bf
  58}, p. 1494  (1987).

\bibitem{alexeyev}
E.~Alexeyev {\em et~al.}, Detection of the neutrino signal from sn 1987a in the
  lmc using the inr baksan underground scintillation telescope, {\em Physics
  Letters B} {\bf 205}, 209   (1988).

\bibitem{superK}
Y.~Fukuda {\em et~al.}, The super-kamiokande detector, {\em Nucl. Instrum.
  Meth.} {\bf A501}, 418  (2003).

\bibitem{lvd_det}
M.~Aglietta {\em et~al.}, The most powerful scintillator supernovae detector:
  Lvd, {\em Il Nuovo Cimento A Series 11} {\bf 105}, 1793  (1992), cited By
  103.

\bibitem{kamland}
F.~Suekane {\em et~al.}, An overview of the kamland 1-kiloton liquid
  scintillator2004.
\newblock \href{https://arxiv.org/abs/physics/0404071}{arXiv:physics/0404071}.

\bibitem{icecube}
M.~Aartsen {\em et~al.}, The icecube neutrino observatory: instrumentation and
  online systems, {\em Journal of Instrumentation} {\bf 12}, P03012 (mar 2017).

\bibitem{Antonioli2004}
P.~Antonioli {\em et~al.}, Snews: the supernova early warning system, {\em New
  Journal of Physics} {\bf 6}, 114 (sep 2004).

\bibitem{Al_Kharusi_2021}
S.~A. Kharusi {\em et~al.}, {SNEWS} 2.0: a next-generation supernova early
  warning system for multi-messenger astronomy, {\em New Journal of Physics}
  {\bf 23}, p. 031201 (mar 2021).

\bibitem{Abbott2017}
B.~P. Abbott {\em et~al.}, Multi-messenger observations of a binary neutron
  star merger*, {\em The Astrophysical Journal} {\bf 848}, p. L12 (oct 2017).

\bibitem{ligo}
J.~Aasi {\em et~al.}, {Advanced LIGO}, {\em Class. Quant. Grav.} {\bf 32}, p.
  074001  (2015).

\bibitem{virgo}
F.~Acernese {\em et~al.}, {Advanced Virgo: a second-generation interferometric
  gravitational wave detector}, {\em Class. Quant. Grav.} {\bf 32}, p. 024001
  (2015).

\bibitem{KAGRA}
T.~Akutsu {\em et~al.}, Overview of \protect{KAGRA}: Detector design and
  construction history  (2020),
  \href{https://arxiv.org/abs/2005.05574}{arXiv:2005.05574}.

\bibitem{Ott_2009}
C.~D. Ott, The gravitational-wave signature of core-collapse supernovae, {\em
  Classical and Quantum Gravity} {\bf 26}, p. 063001 (feb 2009).

\bibitem{Abdikamalov:2020jzn}
E.~Abdikamalov, G.~Pagliaroli and D.~Radice, Gravitational waves from
  core-collapse supernovae  (2020),
  \href{https://arxiv.org/abs/2010.04356}{arXiv:2010.04356}.

\bibitem{powell2020}
J.~Powell and B.~M\"uller, {Three-dimensional core-collapse supernova
  simulations of massive and rotating progenitors}, {\em Mon. Not. Roy. Astron.
  Soc.} {\bf 494}, 4665  (2020).

\bibitem{Szczepanczyk}
M.~Szczepanczyk, J.~Antelis, M.~Benjamin, M.~Cavaglia, D.~Gondek-Rosinska,
  T.~Hansen, S.~Klimenko, M.~Morales, C.~Moreno, S.~Mukherjee, G.~Nurbek,
  J.~Powell, N.~Singh, S.~Sitmukhambetov, P.~Szewczyk, J.~Westhouse, O.~Valdez,
  G.~Vedovato, Y.~Zheng and M.~Zanolin, Detecting and reconstructing
  gravitational waves from the next galactic core-collapse supernova in the
  advanced detector era  (2021),
  \href{https://arxiv.org/abs/2104.06462}{arXiv:2104.06462}.

\bibitem{halimtesi}
O.~Halim, Searching for core-collapse supernovae in the multimessenger era: Low
  energy neutrinos and gravitational waves, PhD thesis, Gran Sasso Science
  Institute (GSSI)2020.

\bibitem{halim2019}
O.~Halim {\em et~al.}, Expanding core-collapse supernova search horizon of
  neutrino detectors, {\em Journal of Physics: Conference Series} {\bf 1468},
  p. 012154 (feb 2020).

\bibitem{Klimenko_2004}
S.~Klimenko and G.~Mitselmakher, A wavelet method for detection of
  gravitational wave bursts, {\em Classical and Quantum Gravity} {\bf 21},
  S1819 (sep 2004).

\bibitem{Klimenko2008}
S.~Klimenko {\em et~al.}, A coherent method for detection of gravitational wave
  bursts, {\em Classical and Quantum Gravity} {\bf 25}, p. 114029 (may 2008).

\bibitem{dragotesi}
M.~Drago, Search for transient gravitational wave signals with unknown waveform
  in the ligo virgo network of interferometric detectors using a fully coherent
  algorithm, PhD thesis, Universit{\`a} degli Studi di Padova2010.

\bibitem{Necula_2012}
V.~Necula, S.~Klimenko and G.~Mitselmakher, Transient analysis with fast
  wilson-daubechies time-frequency transform (jun 2012).

\bibitem{em_gw_ccsne}
B.~P. Abbott {\em et~al.}, Optically targeted search for gravitational waves
  emitted by core-collapse supernovae during the first and second observing
  runs of advanced ligo and advanced virgo, {\em Phys. Rev. D} {\bf 101}, p.
  084002 (Apr 2020).

\bibitem{jankarev}
H.-T. Janka, Explosion mechanisms of core-collapse supernovae, {\em Annual
  Review of Nuclear and Particle Science} {\bf 62}, 407  (2012).

\bibitem{OConnor:2010moj}
E.~O'Connor and C.~D. Ott, {Black Hole Formation in Failing Core-Collapse
  Supernovae}, {\em Astrophys. J.} {\bf 730}, p.~70  (2011).

\bibitem{radice}
D.~Radice, V.~Morozova, A.~Burrows, D.~Vartanyan and H.~Nagakura,
  {Characterizing the Gravitational Wave Signal from Core-Collapse Supernovae},
  {\em Astrophys. J. Lett.} {\bf 876}, p.~L9  (2019).

\bibitem{dimmelmeier2008}
H.~Dimmelmeier {\em et~al.}, Gravitational wave burst signal from core collapse
  of rotating stars, {\em Phys. Rev. D} {\bf 78}, p. 064056 (Sep 2008).

\bibitem{Scheidegger:2010en}
S.~Scheidegger, R.~Kaeppeli, S.~C. Whitehouse, T.~Fischer and M.~Liebendoerfer,
  {The Influence of Model Parameters on the Prediction of Gravitational wave
  Signals from Stellar Core Collapse}, {\em Astron. Astrophys.} {\bf 514}, p.
  A51  (2010).

\bibitem{woosley}
S.~Woosley and A.~Heger, {The Progenitor stars of gamma-ray bursts}, {\em
  Astrophys. J.} {\bf 637}, 914  (2006).

\bibitem{hudepohl}
L.~H{\"u}depohl, Neutrinos from the formation, cooling and black hole collapse
  of neutron stars, PhD thesis, Technische Universit{\"a}t M{\"u}nchen2014.

\bibitem{pagliaroli2009}
G.~Pagliaroli {\em et~al.}, Improved analysis of sn1987a antineutrino events,
  {\em Astroparticle Physics} {\bf 31}, 163   (2009).

\bibitem{pagliaroli_ccsn}
F.~Vissani, G.~Pagliaroli and M.~L. Costantini, {A parameterized model for
  supernova electron antineutrino emission and its applications}, {\em J. Phys.
  Conf. Ser.} {\bf 309}, p. 012025  (2011).

\bibitem{Klimenko:2015ypf}
S.~Klimenko, G.~Vedovato, M.~Drago, F.~Salemi, V.~Tiwari, G.~A. Prodi,
  C.~Lazzaro, K.~Ackley, S.~Tiwari, C.~F. Da~Silva and G.~Mitselmakher, {Method
  for detection and reconstruction of gravitational wave transients with
  networks of advanced detectors}, {\em Phys. Rev. D} {\bf 93}, p. 042004
  (2016).

\bibitem{Drago:2020kic}
M.~Drago, V.~Gayathri, S.~Klimenko, C.~Lazzaro, E.~Milotti, G.~Mitselmakher,
  V.~Necula, B.~O'Brian, G.~A. Prodi, F.~Salemi, M.~Szczepanczyk, S.~Tiwari,
  V.~Tiwari, G.~Vedovato and I.~Yakushin, Coherent waveburst, a pipeline for
  unmodeled gravitational-wave data analysis  (2021),
  \href{https://arxiv.org/abs/2006.12604}{arXiv:2006.12604}.

\bibitem{SNTargeted2016}
B.~P. Abbott {\em et~al.}, First targeted search for gravitational-wave bursts
  from core-collapse supernovae in data of first-generation laser
  interferometer detectors, {\em Phys. Rev. D} {\bf 94}, p. 102001 (Nov 2016).

\bibitem{Abbott:2020qfu}
B.~P. Abbott {\em et~al.}, {Prospects for observing and localizing
  gravitational-wave transients with Advanced LIGO, Advanced Virgo and KAGRA},
  {\em Living Rev. Rel.} {\bf 23}, p.~3  (2020).

\bibitem{TheLIGOScientific:2014jea}
J.~Aasi {\em et~al.}, {Advanced LIGO}, {\em Class. Quant. Grav.} {\bf 32}, p.
  074001  (2015).

\bibitem{TheVirgo:2014hva}
F.~Acernese {\em et~al.}, {Advanced Virgo: a second-generation interferometric
  gravitational wave detector}, {\em Class. Quant. Grav.} {\bf 32}, p. 024001
  (2015).

\bibitem{marektesi}
M.~J. Szczepa\'nczyk, Multimessenger astronomy with gravitational waves from
  core-collapse supernovae, PhD thesis, Embry-Riddle Aeronautical
  University2018.

\bibitem{Agafonova2014}
N.~Agafonova {\em et~al.}, Implication for the core-collapse supernova rate
  from 21 years of data of the large volume detector, {\em Astrophys.\ J.} {\bf
  802}, p.~47  (2015).

\bibitem{Ikeda2007}
M.~Ikeda {\em et~al.}, Search for supernova neutrino bursts at
  super-kamiokande, {\em Astrophys. J.} {\bf 669}, 519  (2007).

\bibitem{Abe2016}
K.~Abe {\em et~al.}, Real-time supernova neutrino burst monitor at
  super-kamiokande, {\em Astropart. Phys.} {\bf 81}, 39  (2016).

\bibitem{Casentini2018}
C.~Casentini {\em et~al.}, Pinpointing astrophysical bursts of low-energy
  neutrinos embedded into the noise, {\em JCAP} {\bf 1808}, p. 010  (2018).

\bibitem{kam_bg}
K.~Eguchi {\em et~al.}, First results from kamland: Evidence for reactor
  antineutrino disappearance, {\em Phys. Rev. Lett.} {\bf 90}, p. 021802 (Jan
  2003).

\end{thebibliography}

\end{document}